\documentstyle[epsf,aps,prl,multicol]{revtex}
\begin{document}
\newcommand{\epsi}{\epsilon}
\newcommand{\eps}{\varepsilon}  
\newcommand{\vare}{\varepsilon}  
\newcommand{\vareps}{\varepsilon}  
\newcommand{\De}{$\Delta$}
\newcommand{\de}{$\delta$}
\newcommand{\mc}{\multicolumn}
\newcommand{\be}{\begin{eqnarray}}
\newcommand{\ee}{\end{eqnarray}}
\newcommand{\einf}{\varepsilon^\infty}
\newcommand{\ez}{\varepsilon^0}  
\newcommand{\nvec}{{\bf \hat{ n}}}  
\newcommand{\Pvec}{{\bf P}}  
\newcommand{\Evec}{{\bf E}}  
\renewcommand{\L}{{\rm L}}
\newcommand{\T}{{\rm T}}
\newcommand{\fine}{ \begin{verbatim} -------- \end{verbatim} }  

\draft \title{\bf 
Polarization fields in nitride  nanostructures: 
theory and practical implications}
\author{Fabio Bernardini and Vincenzo Fiorentini}

\address{Istituto Nazionale per la Fisica della Materia 
and Dipartimento  di Fisica, Universit\`a di 
Cagliari, I-09124 Cagliari, Italy}
\date{31 March 1998}
\maketitle

\begin{abstract}
Huge built-in electric fields 
are predicted to exist in 
wurtzite  III-V nitrides
 thin films and multilayers.
Such fields originate from heterointerface discontinuities 
of the macroscopic 
bulk polarization of the nitrides.
We discuss
 the theoretical background and the intriguing
 practical implications of polarization fields
  for nitride nanostructures.
\end{abstract}

\begin{multicols}{2}
III-V nitrides represent a new frontier of semiconductor physics.
One of their unusual basic properties, macroscopic polarization,
  offers unique opportunities for device design and
 basic investigations.
Polarization   manifests itself as built-in electrostatic fields in
polarized materials interfaced  to different media.
 These fields affect the characteristics,  performance, and
response of multilayer nanostructured devices.
In this Letter  we  discuss the theoretical basis and 
practical 
 implications of polarization fields in nitride nanostructures.
Polarization fields turn out to give  III-V nitrides  a 
considerable potential for novel device design and simulations. 

\par {\it General -- }
The  dipole moment per unit volume
of a finite dielectric is
the {\it longitudinal} polarization $\Pvec_\L$,  also identified with
(minus) the screened field 
generated by the net polarization charge at the sample surfaces.
 $\Pvec_\L$ is  experimentally accessible, but its
direct calculation is impractical.
Recent advances 
\cite{Resta,VKS} have provided a
 route to this quantity  through
 a novel, rigorous definition of the polarization
in a periodic system:
the {\it transverse} polarization $\Pvec_\T$
is the  gauge-invariant Berry's phase
of the Bloch orbitals, accumulated in an adiabatic transformation
of  the system from some reference state to its actual state
\cite{Resta}.
$\Pvec_\T$, which can equivalently be viewed as the integrated
polarization current flowing through the crystal during the
transformation, has no relation with the charge density of
the polarized dielectric. Importantly, $\Pvec_\T$
can now be calculated accurately  from first-principles
density-functional calculations \cite{Resta,conti.pol,noi.pol,noi.prl}.
A key point is that the
{\it absolute} polarization of a material can be obtained with no
arbitrariness by  referencing its polarization to
that of a system for which $\Pvec$=0 by symmetry
or otherwise.  For  wurtzite nitrides \cite{noi.pol}, this may
be vacuum or  the  zincblende phase. 
Finally, 
although  not directly measurable, $\Pvec_\T$ gives access to 
 $\Pvec_\L$  through the classical relation
\be
      \Pvec_\T = \tensor{\eps}\!_0 ~\Pvec_\L,
\ee
once the static dielectric tensor
 $\tensor{\eps}\!_0$ has been measured or computed
 \cite{noi.prl}. 
Here we deal with
the total polarization (either longitudinal or transverse) 
 ${\bf P}={\bf P^{\rm (0)}}$ +
 $\delta {\bf P}$($\epsilon$) in a given
strain state at zero temperature, in the absence of external
fields. We point out that in wurtzite nitrides the total
polarization at zero strain, known as spontaneous, is non-zero and
large.
Unlike the strain-induced piezoelectric terms $\delta {\bf P}$($\epsilon$),
the spontaneous polarization has a fixed direction and magnitude for
any crystal structure.

The polarization  of a material $A$ manifests itself  
 when its bulk  periodicity is broken,  e.g. as a
local charge accumulation at the
 interface with a  different medium $B$. 
Therewith  only polarization differences are
accessed,  as  required by theory. If
the interface is  insulating and gap-states--free,  it can be
proven \cite{VKS} that the  areal interface charge density is 
\be
  \sigma_{\rm int} 
= \pm \nvec \cdot (\Pvec_\L^{A} - \Pvec_\L^{B}),
\label{eq.sigma}
\ee
where  $\nvec$  is the interface normal, and the sign of
$\sigma_{\rm int}$ depends on the relative orientation of the media.
This equality has been verified directly in ab initio calculation
on nitride  interfaces \cite{noi.prl,noi.off}; the 
polarization charge density was found
to be typically $\sim$ 5 $\times$ 10$^{12}$ cm$^{-2}$, 
 localized in an interface region less than $\sim 5$ \AA\, thick.
As a consequence of this charge accumulation  \cite{nota2}, macroscopic
 electrostatic fields (screened by electronic and ionic responses)
 exist  in the interfaced media.  For the moment, we 
 assume that these fields are uniform. Due to Gauss'
law,  upon crossing the interface the field changes by
\be
   \Delta \Evec =  4\pi \nvec \,\sigma_{\rm int}.
\label{delta.E}
\ee
In the absence of external fields,
the values of the fields in the interfaced media are
determined by the bulk polarizations of the layers involved in the 
structure. For example,  the  electrostatic field in a finite,
isolated, and surface-state free slab of a polarized material
is $-4 \pi$ {\bf P}$_{\rm L}$. The field in the
vacuum is zero by construction because of the overall neutrality of
the slab. Similarly, if two slabs of differently polarized materials $A$ 
and $B$ are interfaced to form a   finite and isolated $A$/$B$ slab
the fields will equal  $-4 \pi$ {\bf P}$^{A}_{\rm L}$ in material
$A$ and  $-4 \pi$ {\bf P}$^{B}_{\rm L}$ in material $B$,
and the field  in the vacuum will again be zero.
It is straightforward to show that
this follows directly from the superposition principle of
electrostatics. 
 
We now discuss the  role of polarization fields in
structures  (overlayers,  quantum wells, superlattices) of relevance
to device applications of III-N compounds  and alloys,
starting with the issue of free-carrier screening.

\par {\it Screening -- }
Uniform electrostatic fields, such as those  generated by macroscopic
polarization discontinuities, may not be sustained by an
arbitrarily thick sample \cite{gonze}.  At 
{\it zero temperature} the field-induced 
potential drop across the system will be larger than the band gap
for thickness $l \geq l_c \sim E_{\rm gap}/|\Evec| =
 E_{\rm gap}/4\pi\sigma$.
The metastable state thus realized, whereby the 
valence  edge at one end of the sample is higher than the
conduction edge at the other end,  can reach the
ground state by charge tunneling  across the sample, typically 
 in very long times. 

By contrast (e.g.) {\it  at room temperature},
 the presence of free carriers
forbids the  existence of 
 a non-zero {\it uniform} macroscopic field, as the latter  would 
cause  a permanent current to flow across the sample in the
absence of an external electromotive power. The paradox
is resolved by the very presence of  free carriers, which
 screen the field  away from the  interface. In the small-field 
Thomas-Fermi picture, the field decays  exponentially
with characteristic
lengths of order 0.1 $\mu$m for the nitrides. This is of course
irrelevant for nanostructure layers of $\sim$ 50 \AA\, in size.
However, the fields involved in typical nitride structures 
 are of order 100 MV/m (see below), so that 
the small-field limit is invalid, and the
 full  description of (classical) space charge layers
must be applied \cite{mermin}:
the field is then screened  out
over  typical lengths as small as 
10 \AA\, \cite{ando}. This strong screening
would  prevent the field-induced  potential drop to exceed significantly
k$_{\rm B}$T \cite{mermin}. However, one should be aware
that the  fields in question are also large enough  to require a
quantum treatment of the  space charge distribution; since typical
wavefunctions extend over $\sim$ 60 \AA \cite{ando}, the
 over-exponential \cite{mermin}  damping  of the field may be
 preempted in sufficiently thin layers. While 
quantitative predictions require a
 full quantum treatment for general fields and the specifics
of  device parameters,
 it is sensible to  expect  essentially uniform fields up to layer 
thicknesses of order 100-150 \AA\, in the nitrides. 
In any case, it must be kept in
mind that the   polarization field contributions
to the nanostructure potential
 are {\it fixed} elements of that potential, that may be
 screened to different extents
  by free carriers or otherwise, depending on the 
specific application, geometry, and material.

Carrier screening also preempts dielectric breakdown in
massive samples. This  might be a serious possibility, since
the uniform fields  in the absence of free
carriers  ($\sim$ 500 MV/m)  are much larger than the dielectric
breakdown fields  of high power dielectrics (10-50 MV/m). Of
course, breakdown is not an issue in  thin nanostructures.

\par {\it Overlayers -- }
We now consider a  polarized pseudomorphic overlayer on a
thick heterogeneous substrate, assuming the system 
to be insulating and  surface/interface-states-free. The
layer surface and its interface to the substrate are
 charged, and a surface-normal electrostatic field exists 
inside the overlayer. In the absence of external fields, the
internal field is
\be
    \Evec = 4\pi \nvec \,\sigma^{\rm surf}  = -4\pi \,
(\nvec\cdot\Pvec_\L^{\rm overl})\, \nvec
\label{eq.field}
\ee
with $\nvec$  the surface normal versor. 
 To simplify notation, we henceforth
restrict ourselves to  (0001)-grown nitride layers
where  the total bulk polarization $\Pvec_\L$ is surface-normal,
hence  $(\nvec\cdot\Pvec_\L^{\rm overl})\, \nvec = \Pvec_\L$.

   The result Eq. (\ref{eq.field})
stems again from the superposition principle.
In this configuration, there is a direct proportionality
between longitudinal polarization and electrostatic field in the
layer. In particular, if the overlayer  is unstrained, the
field is proportional to the spontaneous polarization. The latter  has
been calculated \cite{noi.pol} to be negative for all   the III-V
nitrides,  so that $\Evec$ will point in the ($0001$) direction.
According to Eq. (\ref{eq.field}), typical values for the electrostatic
fields in unstrained overlayers range from 250 MV/m,  (InN)
 to 900 MV/m (AlN) (at least in the regime where  screening is not yet
playing a role). Typical polarization charge densities at the relevant
interfaces are in the $10^{12}$ cm$^{-2}$.  
Overlayers strained in the $a-$plane  carry in addition a piezoelectric
polarization along the $c$ axis. Due to the huge piezoelectric
constants of III-V nitrides \cite{noi.pol}, the piezoelectric
and spontaneous  polarizations are generally comparable in magnitude.
Epitaxial strains, depending on their sign and size, may then cause a
strong increase and/or a sign reversal of the total polarization.

As an example of the practical consequences of the above, consider
the band offset at an $A$/$B$ heterojunction,  experimentally determined
via XPS core level alignment as
\be
   \Delta E_v =  [E^A_{cl} - E^A_v]^{\rm bulk} 
              -  [E^B_{cl} - E^B_v]^{\rm bulk} + \Delta E_{cl}^{\rm int},
\label{eq.vbo}
\ee
with $\Delta E_{cl}^{\rm  int}$  the core level binding energy
difference  measured at the interface. 
Since XPS is surface sensitive, this  definition is 
only meaningful if $\Delta E_{cl}^{\rm int}$ is independent 
of the overlayer thickness. 
In a polarized overlayer, the internal electric field will
shift the core binding energies linearly with the
atom's distance from the interface; the XPS signal
 will shift accordingly, as its leading term
originates from the topmost layers.
Because of this shift,   the band 
offset has to be extracted by linear extrapolation to null 
layer thickness of a series of measured $\Delta E_{cl}^{\rm int}$. The
ratio shift/thickness, in turn, gives  directly the screened
polarization field (Eq.(\ref{eq.field})).  

A recent investigation \cite{Rizzi} of band offsets at the strained 
2H-AlN/6H-SiC(0001) interface has indeed revealed this effect. The
extrapolated value of the band offset  agrees with 
ab-initio predictions \cite{Vogl}. 
Using  the theoretical polarization values 
\cite{noi.pol}, we predict a linear
shift of the core levels in the (0001)-oriented \cite{Rizzi} overlayer of 
55 meV/\AA, whereby the spontaneous and piezoelectric 
contributions are about 80 meV/\AA\, and --25 meV/\AA\, respectively.
The experimental shift \cite{Rizzi} 
has the same sign, and a  somewhat smaller magnitude ($\sim$ 30 meV/\AA);
this deviation from the predicted value may be due
to the presence of surface states at AlN (0001)  \cite{DiFelice},
which pin the Fermi level affecting the field-induced 
shift (i.e., surface states partly compensate the effect
of the  $\sim$ 10$^{12}$ cm$^{-2}$ polarization  charge  density,
though of course not the charge itself).
It is  important to note that the neglect of the spontaneous
polarization would lead to a shift {\it  opposite} in sign to that
observed.  

Finally, it is worth mentioning that  the picture does not change  for
polarized overlayers on {\it polarized} substrates. Indeed, as seen
from the overlayer and the vacuum, the substrate is neutral
overall, and produces no uniform field. 

{\it Quantum wells -- } 
While in polarized overlayers the
 field  is proportional
to the overlayer's {\bf P}$_{\rm L}$,
it  easy to prove using the superposition principle of
electrostatics
that 
in an isolated, symmetric (A-B-A) quantum well (QW) the field
generated by interface  charge accumulation at the well's borders is
proportional to the  polarization difference $\Delta \Pvec_\L$ between
cladding and active layers, 
\be
    \Evec_{\rm QW}  = 4\pi \nvec \,\sigma^{\rm int}  = -4\pi \Delta
    \Pvec_\L\,,
\label{eq.qw}
\ee 
while the field outside the  QW is exactly zero.
Within the  QW, the field is effectively
uniform, since typical  screening lenghts are larger than typical 
well thicknesses. 
The assumption of isolated well implies  that 
 the  cladding  layers are sufficiently thick
 in order that influences from  their 
interface to the outer world get screened out; if this were not the
case, free-carriers--screened  field contributions from the far
interfaces would affect the field in the QW. As a consequence of
Eq. (\ref{eq.qw}), if the composition of the cladding layers differs
mildly from that of the well, the internal QW field  may much be
smaller than the absolute polarization value in the material
itself. By the same token, the QW may be made of an {\it  unpolarized}
material and yet have a non-zero internal  field, since the latter is
controlled  only by the polarization {\it difference} with the outer
world. 

The main effect of internal fields in  QWs is to separate spatially
photogenerated or injected carriers of opposite sign. As a
consequence, an increase in well thickness will cause 
increased  recombination times, 
reduced interband transition matrix elements, and 
red-shifted interband transitions. These effects have been studied in
biased unpolarized QWs \cite{Bastard}, and an application to polarized
nitrides appeared recently \cite{Nardelli}. Note that, since the field
is zero outside the well, the  bound states are not metastable, and
 no phenomenon due to finite  escape times
is expected as in classic biased QWs.

An important point about the red shift in QWs is its reversibility.
In  random/ordered/random  GaInP$_2$ alloy  QWs \cite{Ernst},
where built-in polarization fields have been observed,
one finds that applying an appropriate external field
 the transition energy increases, saturating at the value
corresponding to flat-bands conditions in
the active layer: the external bias effectively ``rectifies'' the
polarized well potential. Of course, this effect can be used to
measure the field, hence the polarization, inside the QW.

\par {\it Superlattices -- }
In superlattices,  the field-polarization relation can become highly 
complex. In general, there are no null-field regions, and no simple
proportionality between field and polarization.
The electrostatic field in a superlattice  has  the same period of
the superlattice itself, so that the average electrostatic  field
 $\langle E \rangle$ is null, i.e. there is no uniform field
throughout the whole system. For typical layer thicknesses,
within each homogeneous layer, the field is uniform, and at each interface
one has
\be 
    \Delta \Evec = -4\pi \Delta \Pvec_\L.  
\ee
In the case of alternating
layers of materials $A$ and $B$, of
 dielectric constants $\varepsilon_A$ and  $\varepsilon_B$ and
 respective  thicknesses 
 $l_A$ and $l_B$, 
using Eqs. (\ref{eq.sigma})-(\ref{delta.E}) 
and  periodic boundary conditions, the field in $A$ 
is easily seen to be
\be
    \Evec_A = -4 \pi \, l_B (\Pvec_{\rm T}^{A} - \Pvec_{\rm
 T}^{B})/(l_A \varepsilon_B + l_B \varepsilon_A),
\label{eq:sigma}
\ee
and analogously for $B$.
If the two layers have the same thickness,
\cite{noi.prl} 
\be
    \Evec_A = - 4 \pi \, (\Pvec_{\rm T}^{A} - \Pvec_{\rm T}^{B})/(\eps_A
    + \eps_B),
\label{eq.sl}
\ee
and similarly for $B$.
The presence of these fields, besides its possible practical
 consequences, offers yet another way to measure  the absolute
spontaneous polarization of a given material.
If one of the layers (say, $B$) is made of unpolarized material,
such as GaN in the zincblende structure, Eq. (\ref{eq.sl}) becomes
\be
    \Evec_A = - 4 \pi\, \Pvec_\T^{A} /(\eps_A + \eps_B)
\ee 
whence $\Pvec_\T^A$ is extracted.

{\it Solid solutions -- } 
Solid solutions are ubiquitous in heterostructure
applications. Their polarization is an important parameter
for simulations, and to a first 
approximation it can be  predicted  by 
linear interpolation of the polarizations of bulk III-nitrides
\cite{noi.pol}. 
Assuming that the relevant piezoelectric tensors $\tensor{e}$
    \cite{noi.pol} and strain field $\vec{\epsi}$ are known,
 the transverse polarization of  a pseudomorphically
   strained, e.g., InGaN solid solution is
\be
    \Pvec_\T({\rm In_xGa_{1-x}N}) 
       &\simeq& x ~\Pvec^{\rm (0)}_{\rm InN} + (1-x) ~\Pvec^{\rm
       (0)}_{\rm GaN}
\label{eq.strained}  \\ \nonumber
       &+& \left[ x ~\tensor{e}_{\rm InN} + (1-x) ~\tensor{e}_{\rm
       GaN} \right]  
	    ~\vec{\epsi} \,(x)\, , 
\ee
  containing terms linear as well as quadratic in $x$
(similar relations hold for quaternary solutions). The third term
on the rhs of  Eq. (\ref{eq.strained}) is only present in
pseudomorphic strained growth, and will tend to zero beyond the
critical thickness  at which strain relaxation sets in.

This Vegard-like approach will only yield
 a rough estimate: it is established 
 \cite{Ernst,Froyen} that ordering in cubic III-V solid solutions
can produce  spontaneous polarization  (or change it), 
an effect  not unexpected also in the XN's. Even in the 
random solution,  short-range order in the form of  bond alternation
 may  alter the local electronic structure, hence the  polarization.


{\it Devices: an example -- } Polarization fields offer new possibilities
for device design, and concurrently may affect their performance. 
As a typical example, in 
Fig.1 we sketch  a 
simple near-UV detector composed of a small-$x$, undoped
In$_x$Ga$_{1-x}$N active layer  
cladded by thick $p-$  
and $n-$doped GaN layers.  
As predicted by Eq. \ref{eq.qw}, in
the active quantum well there is  an internal field, which extracts
the photogenerated carriers from the active region and lets them be
easily  collected by a small reverse bias. 
The efficiency of this device should be  very high since
 essentially no tunneling is involved.

\begin{figure}[h]
\unitlength=1cm
\begin{center}
\begin{picture}(5,5)
\put(0.,0)
{\epsfysize=5cm
\epsffile{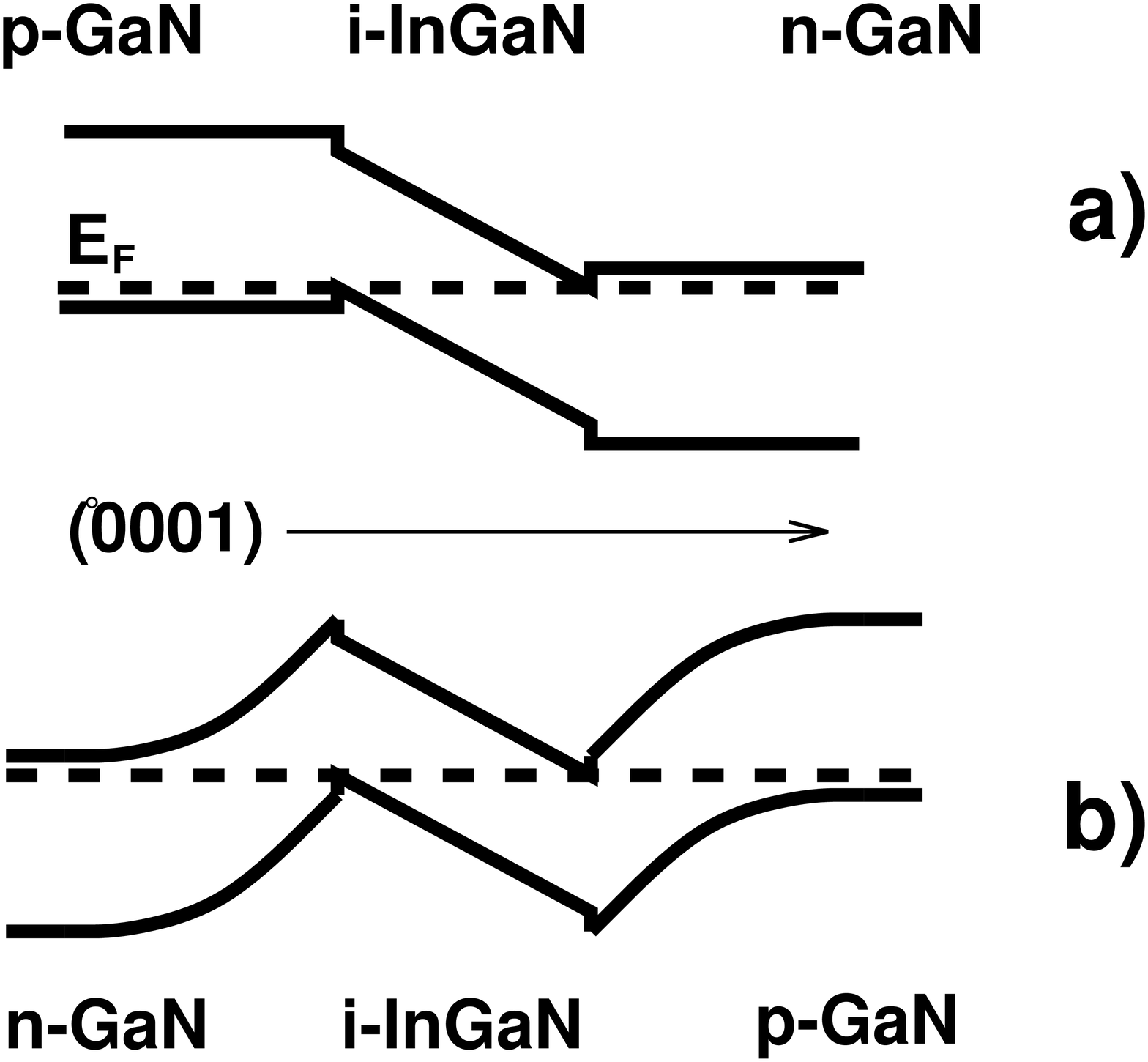}}
\end{picture}
\end{center} 
\centerline{FIG.1\, Electronic potential in the UV detector.} 
\label{fig.1}
\end{figure}

The key issue in this device is the
 crystallographic orientation and ensuing polarity of 
the epitaxially grown  multilayer. 
   Assuming for definiteness $x$=0.1, if the orientation is
 (0001) the correct sequence is $p$-$i$-$n$: 
 the polarization-induced internal field (Eq.(\ref{eq.qw})) 
 in the  strained InGaN well  points in the (000$\bar{1}$)  
  direction, and  provides the desired performance (Fig. 1a). The
 reverse  sequence $n$-$i$-$p$ would  be highly inefficient for this
 specific  application (Fig. 1b). The existence of such  effects of
 polarization fields, which may limit or enhance device performance,
 should be kept in mind in practical work. 

Another potentially useful aspect is that the active layer thickness
may be chosen so that the  potential drop
equals the difference of the Fermi levels of the $n$ and $p$
regions. The   Fermi level is thus equalized across the system
 with flat bands conditions in the doped regions, i.e.  no
 accumulation or depletion layers: assuming as above that
 $x$=0.1, the internal field is 20 mV/\AA, so that to obtain a $\sim$3
eV  potential drop the layer  should be $\sim$ 150 \AA\, thick.
Further,  to avoid strain effects,
 quaternary
 Al$_y$In$_w$Ga$_{1-w-y}$N cladding layers could be used,
with $w$ and $y$  tuned for  lattice matching to In$_x$Ga$_{1-x}$N:
the internal field in the QW  will still be non-zero for any $x$, $y$,
and $w$ since in matching conditions
 Al$_y$In$_w$Ga$_{1-w-y}$N has a  larger spontaneous polarization
 than In$_x$Ga$_{1-x}$N.

In closing, we note that existing
 device realizations, e.g. in the area of optical modulation 
in II-VI nanostructures \cite{mula},
 already exploit built-in piezoelectric
polarization fields. There are, however, at least three
 major elements of novelty in the nitrides:
first, spontaneous polarization, producing fields comparable to, or larger
that the piezoelectric ones; second, the unusual magnitude of the
fields,
typically 2 orders of magnitude larger than in II-VI's;
third, the giant band gap bowing  \cite{bellaiche}
upon alloying  (e.g. of  GaN with 
GaAs), adding further degreees of freedom to nanostructure design.

In summary, we have discussed theory and applications of macroscopic 
polarization concepts to multilayers and devices made of III-V nitride
compounds. We believe that these new concepts will contribute to
open a very fruitful field for device design and  simulation, and 
basic investigations of polarized semiconductors.


\end{multicols}



\begin{thebibliography}{99}
\bibitem{nota2}
Not related to dangling bonds
as in abrupt heterovalent junctions, and  not implying
interface  instability  \cite{noi.off}.
\end{thebibliography}
\end{document}